\documentclass[%
 reprint,
 superscriptaddress,
 amsmath,amssymb,
 aps,
]{revtex4-2}

\usepackage{url}
\usepackage{graphicx}
\usepackage{dcolumn}
\usepackage{bm}
\usepackage{hyperref}

\usepackage{physics}
\begin{document}






\title{Optically-Heralded Entanglement of Superconducting Systems in Quantum Networks}

\author{Stefan Krastanov}
\affiliation{Department of Electrical Engineering and Computer Science, Massachusetts Institute of Technology, Cambridge, MA 02139, USA}
\affiliation{John A. Paulson School of Engineering and Applied Sciences, Harvard University, Cambridge, MA 02138, USA} 

\author{Hamza Raniwala}
\affiliation{Department of Electrical Engineering and Computer Science, Massachusetts Institute of Technology, Cambridge, MA 02139, USA}

\author{Jeffrey Holzgrafe}
\affiliation{John A. Paulson School of Engineering and Applied Sciences, Harvard University, Cambridge, MA 02138, USA} 

\author{Kurt Jacobs} 
\affiliation{U.S. Army Research Laboratory, Computational and Information Sciences Directorate, Adelphi, MD 20783, USA}
\affiliation{Department of Physics, University of Massachusetts at Boston, Boston, MA 02125, USA}

\author{Marko Lončar}
\affiliation{John A. Paulson School of Engineering and Applied Sciences, Harvard University, Cambridge, MA 02138, USA} 

\author{Matthew J. Reagor}
\affiliation{Rigetti Computing, 775 Heinz Ave, Berkeley CA 94710, USA} 

\author{Dirk R. Englund}%
\affiliation{Department of Electrical Engineering and Computer Science, Massachusetts Institute of Technology, Cambridge, MA 02139, USA}\affiliation{Research Laboratory of Electronics, Massachusetts Institute of Technology, Cambridge, MA 02139, USA}

\date{\today}

\begin{abstract}
Networking superconducting quantum computers is a longstanding challenge in quantum science. The typical approach has been to cascade transducers: converting to optical frequencies at the transmitter and to microwave frequencies at the receiver. However, the small microwave-optical coupling and added noise have proven formidable obstacles. Instead, we propose optical networking via heralding end-to-end entanglement with one detected photon and teleportation. This new protocol can be implemented on standard transduction hardware, while providing significant performance improvements over transduction. In contrast to cascaded direct transduction, our scheme absorbs the low optical-microwave coupling efficiency into the heralding step, thus breaking the rate-fidelity trade-off. Moreover, this technique unifies and simplifies entanglement generation between superconducting devices and other physical modalities in quantum networks. 



\end{abstract}

\maketitle

A central challenge in quantum information science is to transfer quantum states between superconducting systems over long distances. The most widely investigated approach is microwave-optical (M-O) quantum state transduction \citep{Stannigel10,safavi2011proposal,Tian12,Clerk12,bochmann2013nanomechanical,bagci2014optical,andrews2014bidirectional,tian2015optoelectromechanical,balram2016coherent,Gard17,higginbotham2018harnessing,covey2019microwave,bartholomew2020chip,welinski2019electron,everts2019microwave,fan2018superconducting,holzgrafeCavity2020,fu2020ground,orcutt2020engineering,hease2020bidirectional,fan2019cavity,fu2020ground,mirhosseini2020quantum,jiang2020efficient}. However, despite concerted effort and tremendous progress in direct M-O transduction, it remains extremely challenging to achieve a high transduction efficiency without adding a significant noise \citep{wu2020microwave}. Moreover, these problems are compounded because a full state transfer between two quantum devices requires sequential M-O and O-M transduction steps \citep{zhong2020proposal}. Here, we propose to replace these M-O-M steps with one round of optically heralded M-M entanglement (requiring the detection of a single optical photon), followed by state teleportation between the quantum devices. In contrast to direct transduction, this photon-heralded entanglement scheme favors low M-O coupling efficiency to eliminate added noise, while assuring on-demand state teleportation by heralding and distilling M-M Bell pairs faster than their decoherence rates. Specifically, for present-day technology, we estimate entanglement rates exceeding 100kHz per channel, an entanglement fidelity exceeding 0.99, and the potential for entanglement purification to reach a fidelity of 0.999 on present-day hardware. Leveraging standard telecom equipment, dense wavelength division multiplexing can boost entanglement rates by an order of magnitude. Our approach also allows efficient and high-fidelity heralded state transfer to other physical modalities including trapped ions, cold atoms, solid-state spin systems, or another traveling photon (corresponding to heralded M-O quantum state transduction). Our approach unifies and simplifies entanglement generation between superconducting devices and other physical modalities in quantum networks. Crucially, our protocol does not require the creation of new hardware: we show that today's transduction hardware can run our heralding scheme while providing orders of magnitude improvement in the networking fidelity compared to the typical deterministic transduction.

\begin{figure}
    \centering
    \includegraphics[width=\columnwidth]{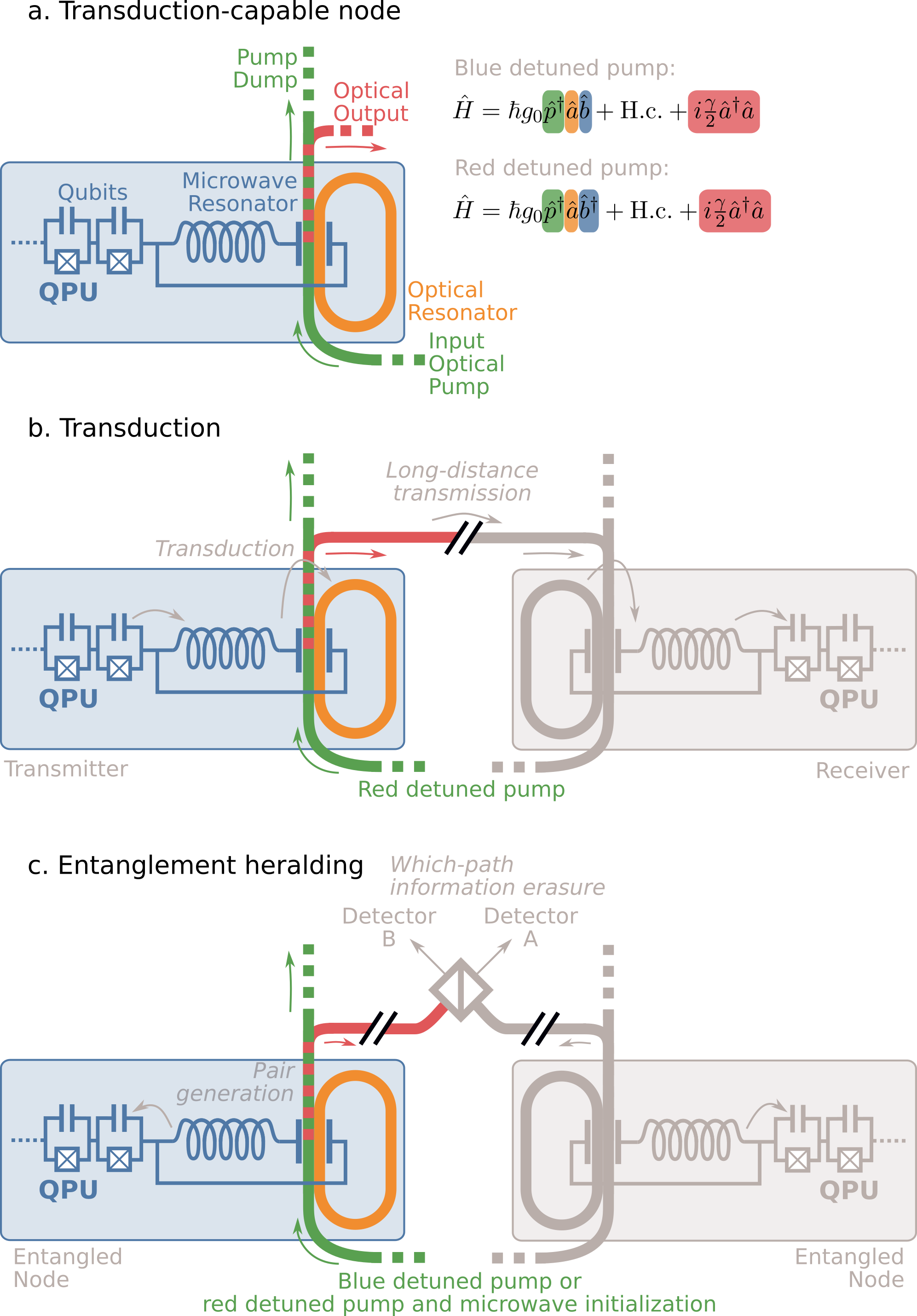}
    \caption{(a) A depiction of a typical electro-optic transducer, using a $\chi^{(2)}$ process whereby a classical pump enables a beam-splitter or two-mode squeezer Hamiltonian between an optical mode and a microwave mode. (b) If the pump is red-detuned from the optical mode it enables a beamsplitter interaction which can transduce a microwave state into an optical state, which is then transmitted over a fiber to a distant node where the same process is used to transduce it back. This is a deterministic, but low-fidelity operation. (c) If the pump is blue-detuned it will create pairs of microwave/optical photons. By detection of the optical photons after erasing the path information we can herald entanglement between the microwave oscillators. This is a high-fidelity, low-efficiency probabilistic operation.}
    \label{fig:drawing}
\end{figure}

Quantum networks were proposed to transfer quantum information between distant cold atoms memories via optical links \citep{cirac1997quantum}, and they now underpin numerous proposed applications \citep{wehner2018quantum}. To connect superconducting systems across such networks, a number of different microwave-to-optical transduction technologies are being investigated \citep{lauk2020perspectives}, including optomechanics \citep{mirhosseini2020quantum,Stannigel10,safavi2011proposal,Tian12,Clerk12,tian2015optoelectromechanical,jiang2020efficient,higginbotham2018harnessing,balram2016coherent,bagci2014optical,andrews2014bidirectional,bochmann2013nanomechanical}, neutral atoms \citep{Gard17, covey2019microwave} and rare-earth-doped crystals  \citep{bartholomew2020chip,welinski2019electron, everts2019microwave}, diamond color centers \citep{neuman2020phononic}, and electro-optic transducers \citep{fan2018superconducting,holzgrafeCavity2020,fu2020ground,orcutt2020engineering,hease2020bidirectional,fan2019cavity,fu2020ground,rueda2019electro,zhu2017preparation, mckennaCryogenic2020, rueda2016efficient}. A critical requirement in these approaches is to optimize conversion efficiency and to minimize added noise, but achieving both simultaneously is extremely challenging.
Recently proposed heralded transduction from a microwave \textit{to an optical photon} by heralding on a microwave Bell pair may help~\citep{zhong2020proposal,zhong2020entanglement}, but two such transduction steps would still be necessary for end-to-end state transfer, entailing high overhead. Alternatively, microwave transduction to proximal diamond color center spins has been proposed for high-speed and high-fidelity quantum network teleportation, but efficient coupling between a single spin and microwave photon has not yet been shown. By contrast, we propose a single-step, optically-heralded scheme for entanglement of two distant superconducting qubits via electro-optic parametric frequency conversion~\citep{mckennaCryogenic2020}, based on the well-known Duan, Lukin, Cirac, and Zoller proposal~\citep{duanLongdistance2001}.

Such a heralded entanglement scheme can be performed using many different MO transduction platforms, but electro-optic transducers are particularly promising for this application. Direct electro-optic transduction 
provides a wide transduction bandwidth (limited only by the microwave lifetime) that enables high-rate entanglement generation even in the low MO coupling regime. Furthermore, recent work has shown that electro-optic transducers can operate with low noise even under optical pumps exceeding $1\, \mathrm{\mu W}$~\citep{hease2020bidirectional, mobassemThermal2020, fu2020ground}. The noise in electro-optic transducers is reduced due to their physical separation of optic and microwave modes, the low thermal resistance provided by their relatively large size and non-suspended structure, and their lack of low-frequency intermediate states. 

A typical electro-optic transducer uses a $\chi^{(2)}$ nonlinear interaction between a classical optical pump mode $\hat{p}$, an optical mode $\hat{a}$, and a microwave mode $\hat{b}$ \citep{rueda2016efficient,holzgrafeCavity2020,fan2018superconducting,fan2019cavity}, as illustrated in Fig.~\ref{fig:drawing}.a. The classical mode is red-detuned with respect to $\hat{a}$ leading to the Hamiltonian 
\begin{align}
  \hat{H} & = \hbar g_0 \hat{p}^\dagger \hat{a}^\dagger \hat{b} + \mathrm{H.c.},
\end{align}
where $g_0$ is the single-photon nonlinear interaction rate.
The $\hat{p}$ mode is in a coherent state with an amplitude high enough that it can be replaced by a classical field with the same amplitude $\hat{p} \rightarrow \sqrt{\langle n_\mathrm{p} \rangle}$, where $\langle n_\mathrm{p} \rangle$ is the average number of photons in the mode. This leads to a beamsplitter-type Hamiltonian $\hat{a}^\dagger\hat{b}+\mathrm{H.c.}$ that can be used for transduction. This transduction is deterministic but its fidelity is low in practice due to the relatively low value of $g_0\sim1\mathrm{kHz}$ compared to the optical loss rates in the system.

If the pump mode $\hat{p}$ is instead blue-detuned with respect to the optical mode $\hat{a}$ the result is two-mode squeezing. This interaction will generate pairs of optical and microwave photons via spontaneous parametric down-conversion (SPDC) \citep{Couteau_2018,guo2017parametric, rueda2016efficient,rueda2020frequency}:
\begin{align}
  \hat{H} & = \hbar g \hat{a} \hat{b} + \mathrm{H.c.} , \\
  g & = g_0 \sqrt{\langle n_\mathrm{p} \rangle}. 
\end{align}
In either case the pump power, $P$, is related to the number of photons in the pump mode as
\begin{equation}
    \langle n_\mathrm{p} \rangle = \frac{4\gamma_\mathrm{e}}{ (\gamma_\mathrm{e}+\gamma_\mathrm{i})^2}\frac{P}{\hbar\omega},
\end{equation}
where $\gamma_{\mathrm{e}}$ is the extrinsic loss rate of the optical mode and $\gamma_\mathrm{i}$ is its intrinsic loss rates. Throughout the majority of this letter we assume the $\gamma_\mathrm{e}$ and $\gamma_\mathrm{i}$ couplings are the same for the $\hat{a}$ and $\hat{p}$ modes.

If we couple the optical mode to a waveguide ending with a photo detector, this SPDC process allows us to herald the production of a single microwave photon by detecting a single optical photon. First we will explore the performance of this procedure, and then we will examine how it enables the heralding of entanglement between two remote microwave systems. Lastly, while the blue-detuned pump is the more natural method for heralded operation, we will also explore an improved way to perform the heralding, which goes back to the use of a red-detuned pump and provides protection against certain errors.

The collapse operator that describes the detection of the optical photon is $\hat{c}=\sqrt{\gamma_\mathrm{e}}\hat{a}$, which leads to a stochastic master equation with non-Hermitian effective Hamiltonian given by~\citep{dum1992monte,molmer1993monte,jacobs2014quantum}
\begin{align}
  \hat{H} & = \hbar g \hat{a} \hat{b} + \mathrm{H.c.} + i\hbar\frac{\gamma_\mathrm{e}}{2}\hat{a}^\dagger\hat{a}.
\end{align}
We will initially assume that the intrinsic loss rate $\gamma_\mathrm{i}$ is negligible, so that all photons are output into the waveguide at rate $\gamma_\mathrm{e}$ (non-zero $\gamma_\mathrm{i}$ will reduce the detection efficiency but not the fidelity). The total loss rate $\gamma=\gamma_\mathrm{i}+\gamma_\mathrm{e}$ is typically much larger than $g$, which simplifies the dynamics. While this hardware constraint leads to limited fidelity even in state of the art transduction devices~\citep{holzgrafeCavity2020}, in the heralding protocol we describe it is essential to obtaining high fidelity. This is because $\gamma\gg g$ is required to ensure that the  SPDC process does not populate the cavity mode with more than one photon at any time. It does, however, limit the rate of photon generation. The lifetime of the microwave oscillator is orders of magnitude longer than the characteristic times of the dynamics studied here and so we take it to be  infinite in our initial analysis~\citep{kjaergaard2020superconducting}. Once the heralded state is prepared in the microwave mode, we swap it out into one of the qubits of the superconducting quantum processing unit (QPU) - a nonlinear operation at which transmon-based devices are becoming very capable. 

We will denote a Fock state with $n_a$ photons in optical mode $\hat{a}$ and $n_b$ in microwave mode $\hat{b}$ as $|n_an_b\rangle$. To obtain a $|11\rangle$ pair on which we can herald the single microwave photon we simply pump the system on the blue-detuned optical mode and let the Hamiltonian evolve, while waiting for a click at the detector. A click heralds the creation of single photon in the microwave mode (a correct approximation as long as $g\ll \gamma_\mathrm{e}$). Solving the dynamics and accounting for the $\frac{\gamma_\mathrm{e}}{\gamma_\mathrm{e}+\gamma_\mathrm{i}}$ drop in efficiency when $\gamma_\mathrm{i}\neq0$ gives a rate of photon generation under a continuous pump of
\begin{align}
  r_0 & = \frac{4g_0^2\langle n_\mathrm{p}\rangle\gamma_\mathrm{e}}{(\gamma_\mathrm{e}+\gamma_\mathrm{i})^2},
\end{align}
which gives $r_0\approx\langle n_\mathrm{p}\rangle10^{-2}\mathrm{Hz}$ at typical $g_0=1\mathrm{kHz}$ and $\gamma_\mathrm{e}=\gamma_\mathrm{i}=100\mathrm{MHz}$.

The easiest way to derive this form for $r_0$ is to restrict oneself to the space spanned by $\{|00\rangle,|11\rangle\}$. For a state $|\psi\rangle = c_0|00\rangle+c_1|11\rangle$ we get the following ODE:
\begin{equation}
\begin{pmatrix}
\dot{c}_0  \\
\dot{c}_1
\end{pmatrix}
=
\begin{pmatrix}
0 & -ig \\
-ig^* & -\frac{\gamma_\mathrm{e}}{2}
\end{pmatrix}
\begin{pmatrix}
c_0  \\
c_1
\end{pmatrix}
\end{equation}
which leads to
\begin{align}\label{eq:ode}
\begin{pmatrix}
c_0  \\
c_1
\end{pmatrix}
&= e^{-\frac{\gamma_\mathrm{e}}{4}t}
\begin{pmatrix}
\frac{\gamma_\mathrm{e}}{4g'}\sinh(g't) + \cosh(g't)  \\
-i\frac{g}{g'}\sinh(g't)
\end{pmatrix},\\
g' & = \sqrt{\frac{\gamma_\mathrm{e}^2}{4}-|g|^2}.
\end{align}
The stochastic master equation tells us that the probability that the photon remains undetected at time $t$ is 
\begin{align}
\langle \psi(t)|\psi(t)\rangle & \sim e^{\left(2g'-\frac{\gamma_\mathrm{e}}{2}\right)t} \sim e^{-\frac{4|g|^2}{\gamma_\mathrm{e}}t},
\end{align}
which indeed corresponds to a Poissonian detection process with rate $r_0$.

\begin{figure}
    \centering
    \includegraphics[width=\columnwidth]{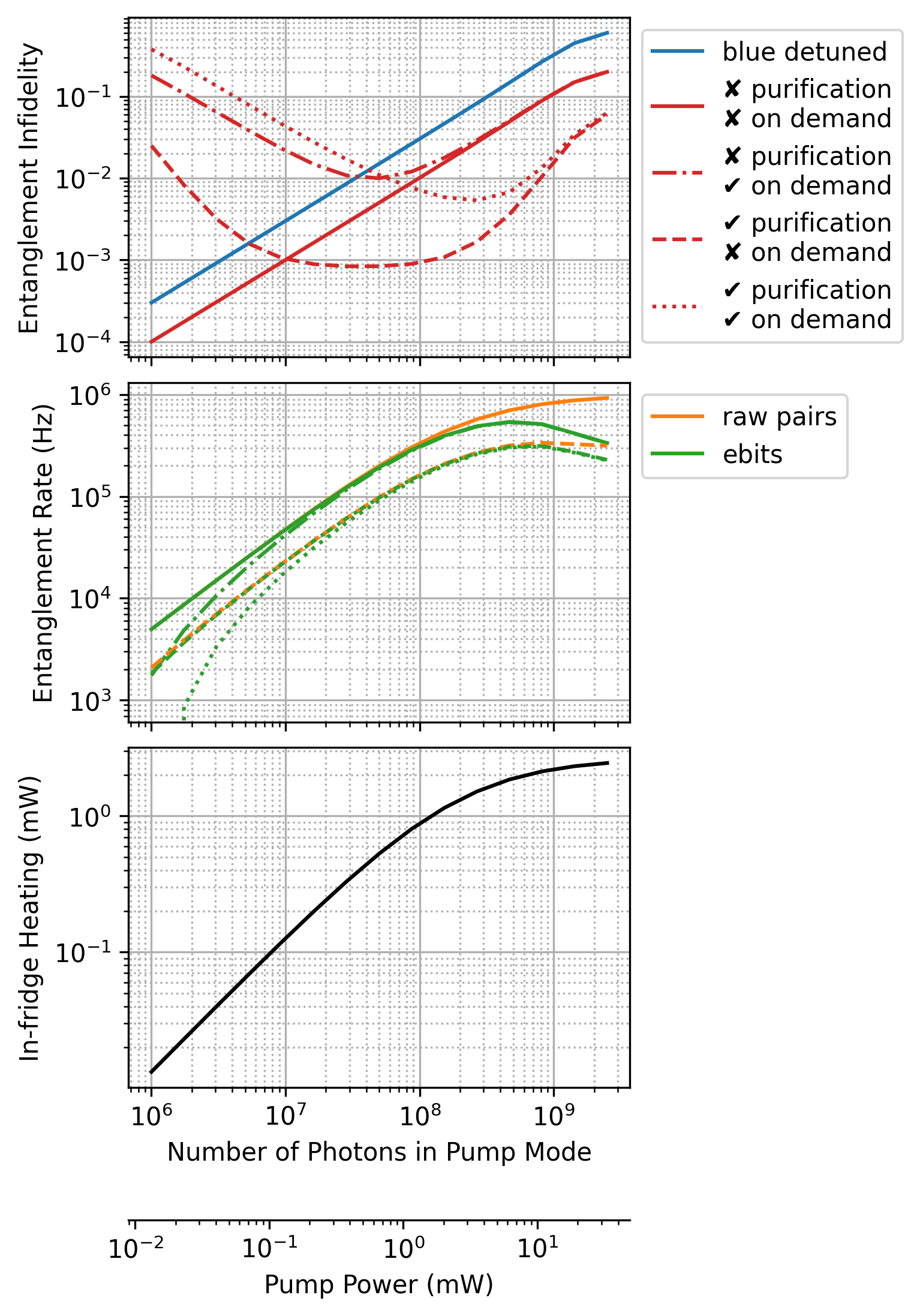}
    \caption{\textbf{In the top figure} we see the infidelity of the obtained microwave Bell pair in different regimes. The blue-detuned ``squeezing'' implementation has higher infidelity due to populating higher-than-one-photon states (blue line). The red-detuned ``beamsplitter'' implementation requires preparing the microwave resonator in the $|1\rangle$ state, but does not suffer from the aforementioned excitations. We further investigate the red-detuned approach followed by purification and/or storage for on-demand use (red lines). Besides the $\sim\frac{g}{\gamma}$ infidelity in the solid lines, we see that storage at low generation-rates causes a infidelity floor, due to finite microwave lifetime. Similarly, purification can be detrimental at low rates due to having to wait for a second pair to be generated. \textbf{The middle figure} takes all red-detuned regimes from the top plot and presents their rates of entanglement generation (orange lines) and rate of equivalent ebit generation, i.e., the hashing yield~\citep{bennett1996mixed} (green lines). We see that purification causes a drop in rate and that the ebit rate suffers at high infidelities. At high rates the curves flatten out due to the finite time for resetting the microwave resonator. \textbf{The bottom figure} gives the in-fridge heating due to the intrinsic loss of the optical resonator. Evaluated at coupling $g_0=1\mathrm{kH}$, optical extrinsic and intrinsic loss $\gamma_\mathrm{e}=\gamma_\mathrm{i}=100\mathrm{MHz}$, microwave loss $\gamma_\mathrm{MW}=1\mathrm{kHz}$, pump wavelength $\lambda_\mathrm{p}=1500\mathrm{nm}$, microwave gate fidelity $0.999$, and microwave resonator reset time of $1\mathrm{\mu s}$, which are state of the art values~\citep{holzgrafeCavity2020,mckennaCryogenic2020, kjaergaard2020superconducting}. Mismatches in the coupling rates of the two nodes, detector dark count rates, and microwave excitations caused by the optical pump are assumed negligible. Errors in the $|1\rangle$ microwave state preparation before heralding is also not included.}
    \label{fig:simulations}
\end{figure}


\begin{figure}
    \centering
    \includegraphics[width=\columnwidth]{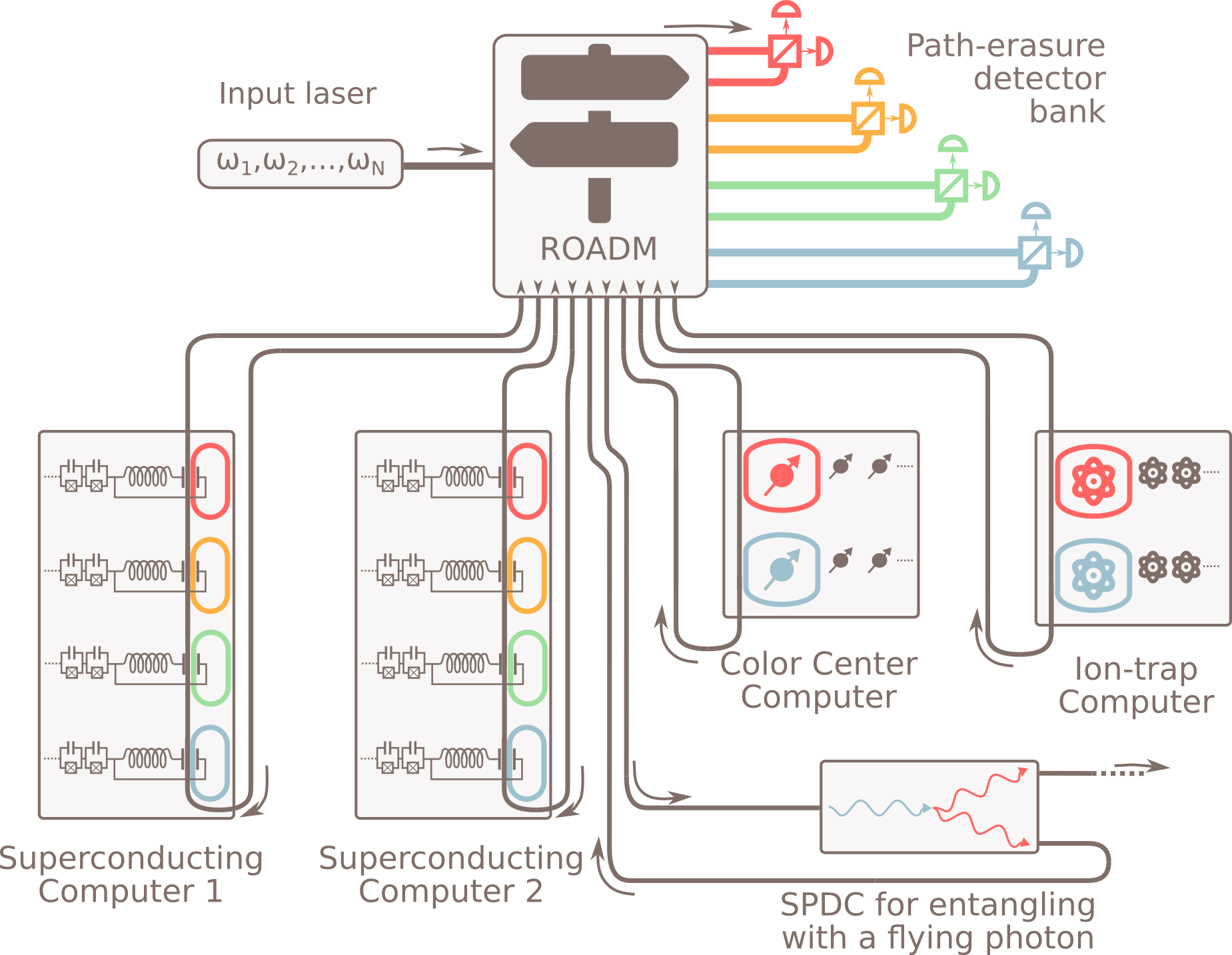}
    \caption{Future outlook for inter-fridge connectivity. Replacing the single path-erasure detection with a reconfigurable optical add-drop multiplexer (ROADM) will allow us to route a frequency comb pump laser to qubits in several quantum nodes. By connecting the ROADM to a bank of path-erasure detectors for each frequency channel, this design will enable multiplexed heralded entanglement generation between multiple fridges of different quantum modalities. Necessary electro-optic modulators and attenuators have been omitted from this schematic.}
    \label{fig:multi}
\end{figure}

If we use the same coherent pump to drive two separate copies of this system and erase the which-path information using a beam splitter (Fig.~\ref{fig:drawing}.c), we will herald the generation of the distributed microwave Bell pair $|01\rangle\pm|10\rangle$ \citep{duanLongdistance2001}.  If the two nodes are mismatched and not calibrated, hence having different interaction rates $g$ and couplings $\gamma_e$, the pair would look like $c_{1\mathrm{l}}|01\rangle\pm c_{1\mathrm{r}}|10\rangle$, where $c_{1\mathrm{l}}$ and $c_{1\mathrm{r}}$ are the coefficients for the left and right network nodes as defined in Eq.~\ref{eq:ode}.

The rate at which these Bell pairs would be generated is 
\begin{equation}
r_\mathrm{e}=2r_0e^{-r_0\Delta t}\frac{\Delta t }{\Delta t + t_\mathrm{r}},    
\end{equation}
 where $r_0\Delta t e^{-r_0\Delta t}$ is the probability for a Poissonian single-click event during the interval $\Delta t$ (the duration of each pump pulse). The factor of $2$ comes from the fact that either of the two nodes can produce a heralded photon. Lastly, the overall rate is lowered, proportional to the duty cycle due to the finite $t_\mathrm{r}$ -- the time necessary for reset of the microwave cavity after each attempt (typically on the order of $1\mathrm{\mu s}$~\citep{kjaergaard2020superconducting}). Throughout the figures in this manuscript we report the maximal value of $r_\mathrm{e}$ after optimizing with respect to $\Delta t$. The probability of more than one event during the interval $\Delta t$ are generally negligible, but lead to infidelities in this protocol at high entanglement rates.
 
The performance of the above method for heralding Bell-pairs is depicted in Fig.~\ref{fig:simulations}. As already discussed, with an internal loss rate $\gamma_\mathrm{i}$ the rate of entanglement generation will be reduced by a factor of $\frac{\gamma_\mathrm{e}}{\gamma_\mathrm{e}+\gamma_\mathrm{i}}$, while the fidelity of the obtained Bell pair is unaffected. Additionally, this heralded scheme ensures the fidelity is unchanged by insertion loss in the optical network (for example due to fiber-transducer coupling losses)~\citep{duanLongdistance2001}, although such losses may require higher in-cryostat optical pump power.  Outside of the regime $g\ll\gamma$ this fidelity would be degraded for two reasons: on one hand, the SPDC process will excite states that contain more than one photon; on the other hand, there will be a small probability, $\frac{c_1^2}{|c_0|^2+|c_1|^2}$, that the SPDC will simultaneously produce a photon in each of the two resonators. Both of these infidelities scale as $\frac{g}{\gamma}$. While the second source of infidelity is unavoidable, the first can be eliminated in the following way: instead of a blue-detuned pump, we can employ the traditional red-detuned pump used in transduction together with a particular state preparation procedure in the microwave hardware. When we reset the microwave cavity we will prepare it in the state $|1\rangle$ instead of the ground state (a high-fidelity operation for modern microwave hardware~\citep{kjaergaard2020superconducting}). This leads to the same ODE as seen in Eq.~\ref{eq:ode}, however the basis for the evolving state is now $|\psi\rangle = c_0|01\rangle+c_1|10\rangle$. Given that this Hamiltonian preserves the total photon number, it cannot excite states outside of the   $\{|01\rangle,|10\rangle\}$ subspace. The same protection against multi-photon states can be achieved by keeping the blue-detuned version of the Hamiltonian and using a strongly anharmonic microwave resonator that suppresses the two-photon excitation. As seen in Fig.~\ref{fig:simulations}, the elimination of unwanted multi-photon states leads to a notable increase in fidelity. Since the two resonators may still generate photons simultaneously, the residual infidelity scales as $\frac{g}{\gamma}$. The blue-detuned approach leads to a microwave entangled pair $|01\rangle\pm|10\rangle$, while in the red-detuned case with state preparation we herald a $|00\rangle\pm|11\rangle$ pair.

We estimate that typical hardware parameters of state-of-the-art devices ($\gamma_\mathrm{e}=\gamma_\mathrm{i}=100\mathrm{MHz}$ and $g_0=1\mathrm{kHz}$ in \citep{holzgrafeCavity2020,kjaergaard2020superconducting}) will allow pair generation rates of $100\mathrm{kHz}$ at fidelities of $0.99$, while suffering $0.1\mathrm{mW}$ of in-fridge heating due to leakage from the pump. This estimate takes into account the finite lifetimes of the optical and microwave cavities. Through simple single-stage purification performed on the microwave superconducting quantum computer the infidelity can be lowered by an order of magnitude while incurring a decrease in the rate of just over a factor of two~\citep{bennett1996mixed}. The fidelity after purification would be chiefly limited by the gate-fidelity of the superconducting quantum computer performing the purification. With gate fidelities of 99\% the benefit of purification is limited as our raw entanglement already reaches that level. In contrast, for gate fidelities of 99.9\% we show in  Fig.~\ref{fig:simulations} that    purification provides a dramatic increase in entanglement fidelity while reducing the entanglement rate only by a factor of approximately two. Lastly, at low entanglement rates the entanglement fidelity reaches a ceiling and starts to worsen because at such low rates the entangled microwave state needs to be stored for a time comparable to the lifetime of the microwave cavity.

The entanglement rate scales as 
$\frac{\gamma_\mathrm{e}^2}{(\gamma_\mathrm{e}+\gamma_\mathrm{i})^4}$,
which is maximal at $\gamma_\mathrm{e}=\gamma_\mathrm{i}$. Therefore, for a fixed pump power and in-fridge heating, the entanglement generation rate is proportional to $\gamma_\mathrm{e}^{-2}$. Modest improvements expected in the next generation of devices can lower the pump power requirements by a factor of 10. Here we have assumed the loss rates are the same for the $\hat{a}$ and $\hat{p}$ modes. At this point the fraction of generated photons that reach the photodetector is only $\frac{\gamma_\mathrm{e}}{\gamma_\mathrm{e}+\gamma_\mathrm{i}}=50\%$. Due to missing half of the heralding events we have to reset the microwave cavity after each attempt, a delay of $\sim1\mathrm{\mu s}$ that limits the maximal rate as seen in the rate plot in Fig.~\ref{fig:simulations}. Alternatively, we can couple the heralding mode $\hat{a}$ strongly to the waive guide, leading to $\gamma_\mathrm{e}\gg\gamma_\mathrm{i}$ for $\hat{a}$, while $\gamma_\mathrm{e}=\gamma_\mathrm{i}$ for the pump mode $\hat{p}$. This would remove the need to reset the microwave resonator, but it would also significantly lower the entangling rate, and is only applicable in the blue-detuned version of our protocol.

In summary, at high entanglement rates (high pump powers) the fidelity of entanglement suffers due to undesired excitations in the non-heralded cavity, while at low rates the fidelity suffers due to the need to store the entangled pair for a long time in the microwave resonator. If we want higher rates of entanglement generation we can counteract the drop in fidelity by performing purification.
Longer term, orders of magnitude improvement in entanglement rate or power requirements are possible through improvements in the resonator quality factor, thanks to the quadratic dependence on $\gamma_\mathrm{e}$.

Several other practical effects can reduce the fidelity of the resulting microwave Bell pairs. These effects include pump-induced heating of the microwave mode, transmission losses between the transducer and the QPU, and false heralding events.

Optical pump light can produce microwave noise by direct absorption in the microwave resonator or heating the transducer node -- a key challenge for all MO transducer platforms. Several techniques can be used to reduce the effects thermal noise in our proposed system, including light shielding \citep{kreikebaumOptimization2016} and immersion in a liquid helium coolant \citep{fan2018superconducting}. Additionally, the microwave-optical transducer can be physically connected to a $1\mathrm{K}$ stage, while radiatively overcoupling the microwave cavity mode to a  thermal bath at $10\mathrm{mK}$, as proposed in previous works \citep{zhong2020proposal,zhong2020entanglement,xu2020radiative}. This allows the transducer to use the greater cooling power of a $1\mathrm{K}$ stage  while the noise is dominated by that of the colder stage. Alternatively, larger-scale custom cryostats for low-background experiments~\citep{singh2016cuore,davis_2020} have shown the requisite cooling power at $100\mathrm{mK}$.

System-integration constraints -- such as the desire to operate the QPU with minimal pump-light interference -- could require physical separation within the cryostat between the transducer node we describe here and the main QPU. In this case, an additional pitch-and-catch transfer ~\citep{campagne2018deterministic,wenner2014catching,axline2018demand,kurpiers2018deterministic} of the microwave Bell state would be required. Losses during this transfer would reduce fidelity, but recent work on high-efficiency pitch-and-catch schemes ~\citep{campagne2018deterministic,wenner2014catching,axline2018demand,kurpiers2018deterministic} suggests this loss can be made small.

False heralding events can occur due to detector dark counts or pump light that leaks to the detectors. Infidelity due to dark counts is likely to be orders of magnitude lower than other errors, with dark count rates on modern photon detectors much smaller than $1\mathrm{kHz}$~\citep{marsili2013detecting,dauler2014review}.
Pump leakage can be minimized through filtering~\citep{piekarek2017high}, which can be done with minimal optical loss by integrating electro-optic filter banks onto the transducer chip \citep{wang2019monolithic}. Appendix B describes one possible on-chip filter design with just 0.5dB insertion loss. 

Mismatches in the coupling rates between the two nodes would lead to a coherent error in which the heralded Bell pair is not a perfectly equal superposition of $|01\rangle$ and $|10\rangle$, thus a high fidelity hardware implementation will require the capability of in-situ calibration of the coupling rates.

The scheme described here extends naturally to heralding entanglement between a diverse set of quantum devices. On one hand, we can multiplex the entanglement heralding over multiple frequency channels, thus improving the entanglement generation rate. We can also use the method to entangle different quantum modalities (see Fig.~\ref{fig:multi}). For example, entangling a superconducting device (where the pump, under a $\chi^{(2)}$ interaction, creates a state $|00\rangle+\varepsilon|11\rangle$) and a trapped ion device (where a conditional reflection of an attenuated pump from a $|+\rangle$ ion state creates a state $|0+\rangle+\varepsilon|1-\rangle$) in which after path erasure we obtain the entangled microwave-ion pair $|0+\rangle\pm|1-\rangle$.
We can also herald the entanglement of the microwave cavity with a flying photon: one of the nodes is of the architecture discussed up to here, while the other node uses the pump in a SPDC photon pair generation experiment calibrated to have the same 
generation rate. In other words one of the nodes can employ a microwave-optical $\chi^{(2)}$ process while the other node employs a purely optical $\chi^{(2)}$ process, leading to heralding the entanglement of a microwave qubit and a flying optical qubit (in the single-rail basis).

As mentioned earlier in this letter, the quantum networking community has explored transduction through heralding over optical-microwave two-mode-squeezing and entanglement before~\citep{zhong2020entanglement,zhong2020proposal,barzanjeh2012reversible,rueda2019electro}. However, these schemes generally provide for the transduction or teleportation of a microwave state into an optical state or the same in the reverse direction. In our approach, while the hardware is virtually the same, the quantum information of interest is never truly carried by an optical mode, rather the optics is used only for the direct entanglement of two microwave modes.


In conclusion, we have proposed a method of heralded entanglement generation between remote microwave photons. Our scheme relies on SPDC using previously demonstrated electro-optic transducers tuned to the low-coupling regime, such that low-efficiency generation of a microwave- and optical-photon pair is the primary transduction process. We have shown that, by heralding the optical photons from two QPUs, we can entangle the accompanying microwave photons remaining in each microwave cavity. Hence, our scheme allows for high-fidelity entanglement in the same hardware used today for low-fidelity transduction. We further demonstrate how to scale our proposed architecture to connect multiple QPUs across several quantum modalities. We believe this entanglement procedure will be valuable for the on-demand, multiplexed entanglement necessary for a quantum network.

\begin{acknowledgements}
We thank Liang Jiang, Marc Davis, Saikat Guha, Michael Fanto, Matthew LaHaye, and Lamia Ateshian for helpful discussions. The Python and Qutip open source communities provided invaluable research software. SK and HR are grateful for the funding provided by the MITRE Quantum Moonshot Program. KJ acknowledges support from a DEVCOM Army Research Laboratory ECI grant. HR acknowledges support from the NSF Center for Ultracold Atoms. JH, ML, and MR acknowledge support from AFRL. JH, DE, and ML acknowledge support from NSF.  
\end{acknowledgements}

\bibliography{bib}

\clearpage
\onecolumngrid 
\appendix

\section{Purification and Ebits}

The state generated by our heralding process would not be a pure entangled state. Rather, it would be of the form
\begin{align}
    \rho & = (1-\varepsilon)|A\rangle\langle A| + \varepsilon|B\rangle\langle B| + \varepsilon|C\rangle\langle C| \\
    |A\rangle & = |01\rangle \pm |10\rangle \\
    |B\rangle & = |00\rangle + |11\rangle \\
    |C\rangle & = |00\rangle - |11\rangle,
\end{align}
where $|A\rangle$ is one of the two possible desired entangled states and $\varepsilon$ is a measure of the infidelity. In the main text we suggest the possibility of performing a purification procedure in order to lower the $\varepsilon$. Notice that the error is not pure depolarization, rather it is biased, leading to the better performance of specific purification protocols. Taking into account the back-action errors introduced by imperfect gates in the purification circuit further complicates the design~\citep{krastanov2019optimized}. However, if the initial fidelity is already high ($\sim99\%$) a simple circuit suffices: quadratic suppression is possible with the prototypical minimal purification circuit, a bilateral CNOT operation between the pair to be purified and a sacrificial pair followed by a Bell measurement on the sacrificial pair. Due to the sacrifice, the effective rate is halved.

We also provide a more abstract effective entangled bits measure, representing the highest amount of perfect Bell pairs distillable from the available imperfect pairs. This distillation procedure assumes access to perfect multi-qubit gates and measurements. Its yield is obtained through the entanglement entropy of the available states, namely
\begin{equation}
    \mathrm{yield} = 1-S(\varepsilon) = 1 - (1-\varepsilon)\log(1-\varepsilon) - 2\varepsilon\log(\varepsilon).
\end{equation}

\section{Pump Filtering}

\begin{figure}[htb]
\centering
	\includegraphics[width=0.5\linewidth]{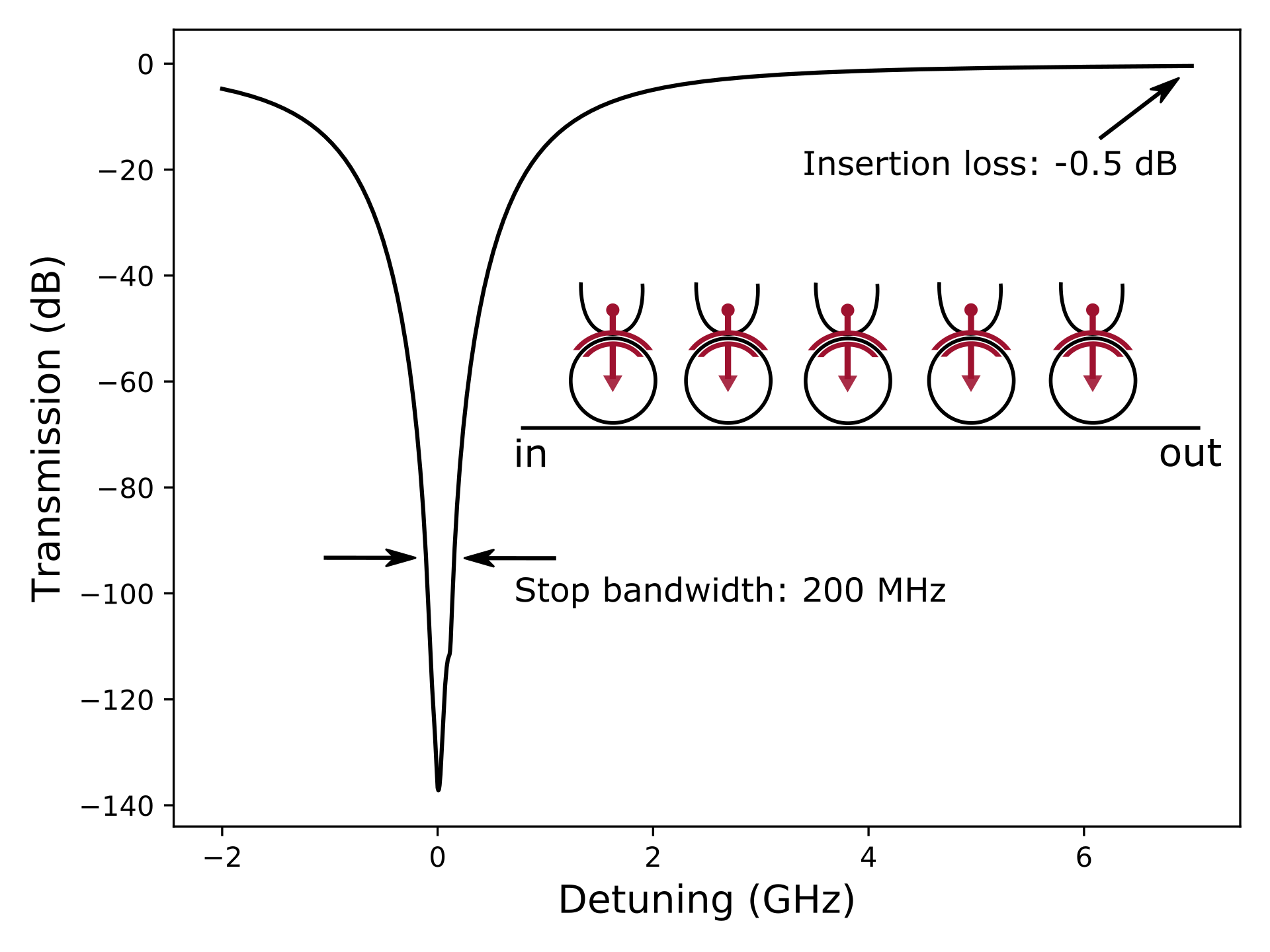}
	\caption{On-chip integrated filter array based on tunable add/drop ring resonators. Simulated transmission assumes intrinsic loss rates of each resonators of $\kappa_i/2\pi = 70 \, \mathrm{Hz}$, and extrinsic coupling to both waveguides of $\overline{\kappa_e/2\pi} = 1 \, \mathrm{GHz}$ with a standard deviation $\Delta \kappa_e/2\pi = 0.1 \, \mathrm{GHz}$ to include typical fabrication uncertainties. Resonance frequencies are also assumed to have standard deviation $\Delta f_0 = 50 \, \mathrm{MHz}$ to include the effects of imperfect voltage tuning. Inset: schematic of the filter array, showing the tuning capacitors (red) and optical waveguides (black).}
	\label{fig:filter}
\end{figure}

The dark counts due to crosstalk from the pump can be minimized by integrating low-loss on-chip pump filters with the transducer~\citep{wang2019monolithic, piekarek2017high}. Filters sufficient for our application have not yet been demonstrated in thin-film lithium niobate, but the low loss rates of ring resonators in this platform makes high-extinction low-loss filtering relatively straightforward. Fig.~\ref{fig:filter} shows one possible filter design with $N_{filter} = 5$ tunable ring add/drop filters. The required $-95 \,\mathrm{dB}$ extinction can be achieved over a stop bandwidth of $\mathrm{SBW} = 200\, \mathrm{MHz}$, with just $-0.5 \, \mathrm{dB}$ of insertion loss at a signal frequency $7\, \mathrm{GHz}$ from the pump frequency. The independent electro-optic tunability of the individual resonators ensures that small fabrication-induced variations in the resonant frequencies can be tuned away without additional heat load to cryostat. 

\section{Purification Results Depending on Gate Fidelity}

\begin{figure}[htb]
\centering
	\includegraphics[width=0.55\linewidth]{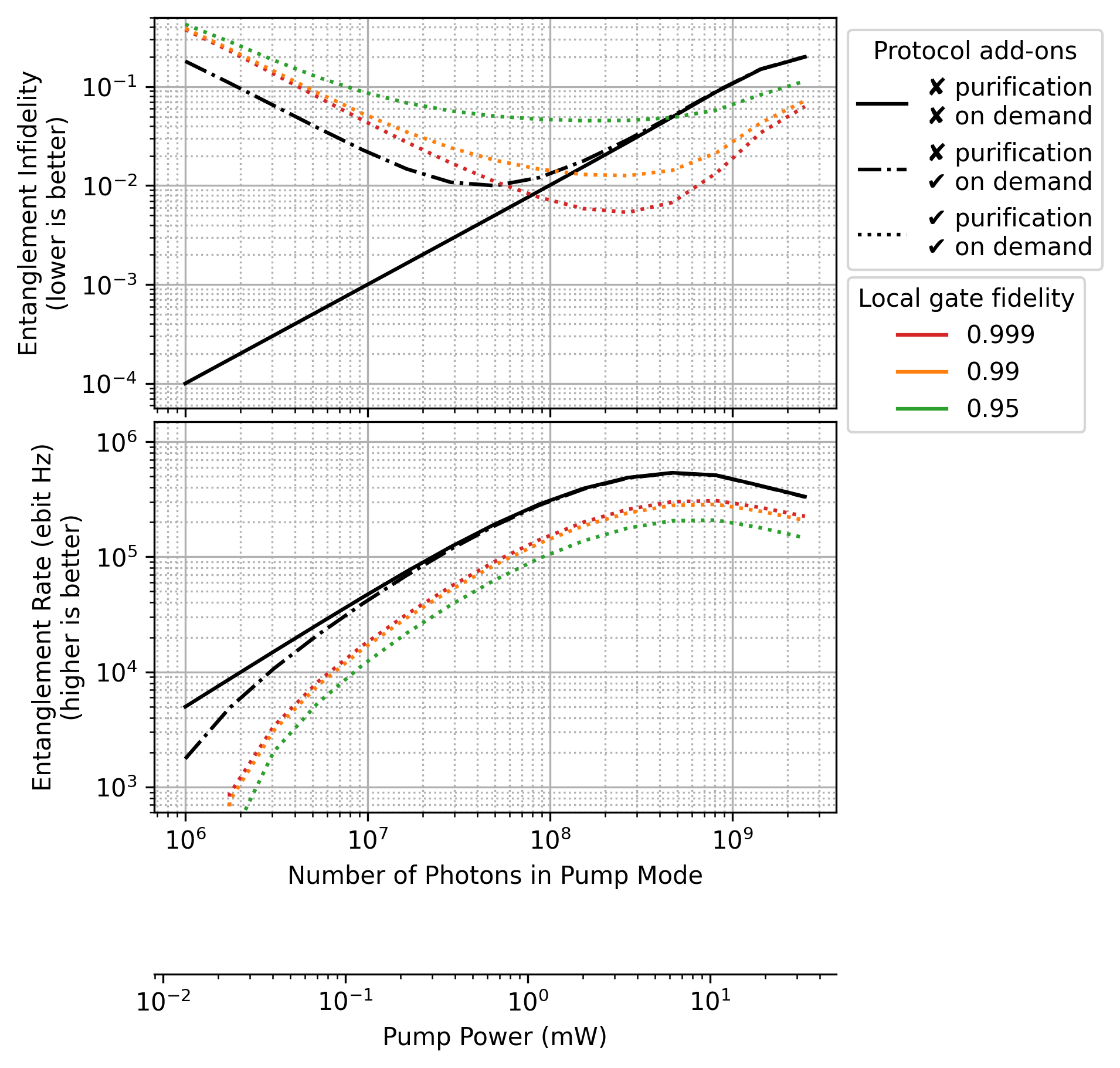}
	\caption{Fidelity and entanglement rate after purification for various local gate fidelities. The black curves correspond to protocol variants in which no purification is performed: immediate use of the Bell pair (solid) and use of the Bell pair after worst-case waiting period (dash-dotted). In color are the fidelities after purification is performed, estimated for various local gate fidelities. Under gate fidelity of 0.99, purification becomes detrimental. The entanglement rate in the bottom subplot is the effective entangled-bit rate, which quantifies not only the rate at which Bell pairs are produced, but also their fidelity.}
	\label{fig:varied_gatefid}
\end{figure}

In the main text we discussed introduced possible "add-ons" to the heralded entanglement protocol. Namely, the possibility of performing simple purification on the entangled states, as well as the effects of memory errors when the Bell pairs are to be stored for on-demand use. In Fig.~\ref{fig:varied_gatefid} we explore the tradeoffs of purification depending on the fidelity of the local gates available at each node. Under gate fidelity of 0.99, purification becomes detrimental.

\end{document}